\documentstyle[12pt,epsfig]{article}
\newcommand{\be}[1]{\begin{equation} \label{(#1)}}
\newcommand{\ee}{\end{equation}}
\newcommand{\ba}[1]{\begin{eqnarray} \label{(#1)}}
\newcommand{\ea}{\end{eqnarray}}
\newcommand{\nn}{\nonumber}
\newcommand{\rf}[1]{(\ref{(#1)})}
\def \lsim {\mbox{${}^< \hspace*{-7pt} _\sim$}}
\def \gsim {\mbox{${}^> \hspace*{-7pt} _\sim$}}

\def\Lv{$L\hspace{-0.5em}/\ $}

\def\znbb{0\nu\beta\beta}
\def\kd{{\rm K}^+\rightarrow \mu^+\mu^+\pi^-}

\def \lg  {\langle}
\def \rg  {\rangle}

\def \znbb {0\nu\beta\beta}
\begin{document}
\begin{center}
   {\Large\bf Lepton number violating processes and Majorana neutrinos\footnote{Talk presented
   by S.Kovalenko at the {\it International Workshop on Neutrino Physics (NANAPino)},
    JINR, Dubna, RUSSIA, July 19-22, 2000}
}\\[3mm]
Claudio Dib, Vladimir Gribanov$\ ^2$, Sergey Kovalenko
\footnote{On leave from the Joint Institute for Nuclear Research, Dubna, Russia}
and  Ivan Schmidt\\[1mm]
{\it Departamento de F\'\i sica, Universidad
T\'ecnica Federico Santa Mar\'\i a, Casilla 110-V, Valpara\'\i so, Chile}
\end{center}
\bigskip
\begin{abstract}
We discuss some generic properties of lepton number violating (\Lv)
processes and their relation to different entries of the Majorana neutrino
mass matrix.  Present and near future experiments searching for these processes, 
except the neutrinoless double beta decay, are unable to probe light(eV mass region) 
and heavy(hundred GeV mass region) neutrinos. On the other hand due to the effect of 
a resonant enhancement, some of \Lv decays can be very sensitive to the intermediate 
mass neutrinos with typical masses in hundred MeV region. These neutrinos may appear 
as admixtures of the three active and an arbitrary number of sterile neutrino species. 
We analyze the experimental constraints on these massive neutrino states and discuss 
their possible cosmological and astrophysical implications.
\end{abstract}
\vskip 0.5cm
\newpage
\section{Introduction}

Recent evidences for neutrino oscillations leave nearly
no room for doubts that neutrinos are massive particles. After all
this point of view is becoming conventional. Solar neutrino
deficit, atmospheric neutrino anomaly and results of LSND neutrino
oscillation experiment all can be explained in terms of neutrino
oscillations implying non-zero neutrino masses and mixings
\cite{nu-oscill}. On the theoretical side \cite{nu-mass} there are
also many indications in favour of non-zero neutrino masses
following from almost all phenomenologically viable models of the
physics beyond the standard model(SM). These models typically
predict Majorana type neutrino masses suggesting that neutrinos
are truly neutral particles.

The neutrino oscillation searches fix both the neutrino mass square difference
$\delta m^2_{ij}=m^2_i-m^2_j$ and the neutrino mixing angles,
leaving the overall mass scale and the CP-phases arbitrary.
Since the latter has no effect on neutrino oscillations, the important
question of whether neutrinos are Majorana or Dirac particles cannot
be answered by these searches.

Majorana masses violate total lepton number conservation
by two units $\Delta L = 2$. Thus lepton number violating(\Lv)
processes represent a most appropriate tool to address the question of
the Majorana nature of neutrinos. A celebrated example of \Lv-process,
most advanced experimentally and theoretically, is the neutrinoless
nuclear double beta ($\znbb$) decay
(for a review see \cite{znbb-rev1,znbb-rev2}).
The $\znbb$-experiments achieved unprecedented sensitivity to
the so called effective Majorana neutrino mass
$\lg m_{\nu}\rg_{ee}$ \cite{znbb-exp}, which in the presence of only light neutrinos
coincides with the entry of the Majorana neutrino 
mass matrix $\lg m_{\nu}\rg_{ee} = M^{(\nu)}_{ee}$.
One may hope to infer information on the other entries from the other
\Lv processes. Many of them have been studied in the literature
in this respect from both theoretical and experimental sides. Among
them there are the decay $\kd$ \cite{Kdec1,Kdec2,Kdec3,LS:2000,Kdec4,GDKS:07.2000},
the nuclear muon to positron \cite{DKT}  or to antimuon \cite{MMM} conversion,
tri-muonium production in neutrino muon scattering \cite{tri-mu}, and the process
$e^+p\rightarrow \bar\nu l^+_1l^+_2 X$ relevant for HERA \cite{HERA},
as well as direct production of heavy Majorana neutrinos at various colliders
\cite{heavy-Majorana}.
Unfortunately sensitivities of the current experiments
searching for these processes are much less than in case of
$\znbb$-decay.
The analysis made in the literature \cite{Kdec3,lnv} leads
to the conclusion that if these processes are mediated by Majorana
neutrino exchange then, except $\znbb$-decay, they can hardly
be observed experimentally. This analysis relies on the current
neutrino oscillation data, and on certain assumptions related to the neutrino
mass matrix.
We will show that, despite the above conclusion being true
for contributions of the neutrino states much lighter or much heavier
than the typical energy of a certain \Lv process, there are still
special windows in the neutrino sector which can be efficiently
probed by searching for some of these processes. In the case of $\kd$
decay this window lies in the neutrino mass range $245\mbox{
MeV}\leq m_{\nu_j}\leq 389$ MeV, where the s-channel neutrino
contribution to the $\kd$ decay is resonantly enhanced, therefore
making this decay very sensitive to neutrinos in this mass domain.
If neutrinos with these masses exist, then from the present
experimental data we can extract stringent limits on their mixing
with $\nu_{\mu}$.

Recently some phenomenological, cosmological and astrophysical issues of
the intermediate mass neutrinos in the MeV mass region have been addressed 
\cite{Dolgov}. This was stimulated by the attempts of explanation of
the KARMEN anomaly in terms of these massive neutrino states \cite{KARMEN}.
Although recent data of the KARMEN collaboration \cite{KARMEN2} have not confirmed
this anomaly, the possible existence of these massive neutrinos remains open,
motivated by the idea of sterile neutrinos $\nu_s$ required for the explanation
of all the neutrino oscillation data including the LSND results. The sterile species $\nu_s$
may mix with the active ones $\nu_{e,\mu,\tau}$ to form massive states with {\it a priori} 
arbitrary masses.  Their existence is the subject of experimental searches as well 
as cosmological and astrophysical constraints.

The paper is organized as follows. In section 2 we discuss a model with sterile
neutrinos and possible spectrum of massive neutrino states. Section 3 is devoted
to some general features of constraints on the neutrino mass matrix, derivable from
the \Lv processes. In section 4 we give the theoretical framework for
$\kd$ decay, and then in section 5 discuss expected rates of this and other \Lv
processes in the light of the present neutrino observations. In section 6 we study
the possible contribution of massive neutrinos in the resonant domain of
the $\kd$ decay and derive the constraints
on the mixing of these neutrinos with $\nu_{\mu}$. Astrophysical and cosmological
implications of these massive hundred MeV neutrinos are shortly addressed.

\section{Majorana neutrino mass matrix and neutrino counting experiments.}

Consider an extension of the SM with the three left-handed weak
doublet neutrinos $\nu'_{Li} = (\nu'_{Le},\nu'_{L\mu},\nu'_{L\tau})$
and n species of the SM singlet right-handed neutrinos
$\nu'_{Ri}=(\nu'_{R1},...\nu'_{Rn})$.
The general mass term for this set of fields can be written as
\ba{mass-term}
\nn
&-& \frac{1}{2} \overline{\nu^{\prime}} {\cal M}^{(\nu)} \nu^{\prime c} +
\mbox{H.c.} =
- \frac{1}{2}
(\bar\nu'_{_L},  \overline{\nu_{_R}^{\prime c}})
\left(\begin{array}{cc}
{\cal M}_L & {\cal M}_D \\
{\cal M}^T_D  & {\cal M}_R \end{array}\right)
\left(\begin{array}{c}
\nu_{_L}^{\prime c} \\
\nu'_{_R}\end{array}\right) + \mbox{H.c.} =\\
&-&\frac{1}{2} \sum_{i=1}^{3+n} m_{\nu i} \overline{\nu^c}_{i}\nu_i
+ \mbox{H.c.}
\ea
Here ${\cal M}_L, {\cal M}_R$ are $3\times 3$ and $n\times n$ symmetric
Majorana mass matrices, ${M}_D$ is $3\times n$ Dirac type matrix.
Rotating the neutrino mass matrix by the unitary transformation
\ba{rotation}
U^T {\cal M}^{(\nu)}U = Diag\{m_{\nu i}\}
\ea
to the diagonal form we end up with $n+3$ Majorana neutrinos
$\nu_i =  U^*_{ki} \nu'_{k}$ with the masses $m_{\nu i}$.
In special cases there may
appear among them pairs with masses degenerate in
absolute values. Each of these pairs can be collected into a Dirac neutrino
field. This situation corresponds to conservation of certain lepton numbers
assigned to these Dirac fields.

The considered generic model must contain at least three observable light neutrinos
while the other states may be of arbitrary mass.  In particular, they may include
hundred MeV neutrinos, which we will consider in section 6. Presence or absence of
these neutrino states is a question for experimental searches.

Let us point out that the presence of more than three light neutrinos
are not excluded by the neutrino counting LEP experiments measuring
the invisible Z-boson width $\Gamma_{inv}$. Actually it
counts not the number of light neutrinos but the number of active flavours.
To see this let us write down the $Z \nu \nu$ interaction term as
\ba{Z-width}
Z^{\mu} \sum_{\alpha = e,\mu,\tau}
\overline{\nu'_{\alpha}} \gamma_{\mu} \nu'_{\alpha} =
Z^{\mu} \sum_{\alpha = e,\mu,\tau}\sum_{m,n=1}^{n+3}
U_{\alpha m}U^*_{\alpha n}
\overline{\nu_{n}} \gamma_{\mu} \nu_{m} \equiv
\sum_{m,n=1}^{n+3} {\cal P}_{mn}
\overline{\nu_{n}} \gamma_{\mu} \nu_{m},
\ea
where the last two expressions are written in the mass eigenstate basis. For the case
of only three massive neutrinos one has  ${\cal P}_{mn}=\delta_{mn}$
as a consequence of unitarity of $U_{\alpha n}$. In general
${\cal P}_{mn}$ is not a diagonal matrix and
flavor changing neutral currents in the neutrino sector become possible at tree level.
However if all the neutrinos are significantly lighter than Z-boson
with the mass $M_Z$ their contribution to the invisible the Z-boson width is
\ba{inv-width}
\Gamma_{inv} =  \sum_{m,n=1}^{n+3} \left |{\cal P}_{mn}\right |^2
\Gamma_{\nu}^{SM} = \Gamma_{\nu}^{SM} \sum_{\alpha,\beta = e,\mu,\tau}
\delta_{\alpha\beta}\delta_{\alpha\beta} =  3\Gamma_{\nu}^{SM},
\ea
where $\Gamma_{\nu}^{SM}$ is the SM prediction for the partial Z-decay width
to one pair of light neutrinos.
This chain of equalities follows again from the unitarity of $U_{\alpha n}$.
Thus, independently of the number of light neutrinos with masses
$m_{\nu}<< M_Z/2$ the factor 3 in the last step counts the number of
weak doublet neutrinos. This conclusion is changed in the presence of
heavy neutrinos $N$ with masses $M_N > M_Z/2$ which do not contribute to
$\Gamma_{inv}$. In this case the  unitarity condition is no longer valid
and the factor 3 is changed to a smaller value.

Having these arguments in mind we introduce in section 6
neutrino states with masses in the hundred MeV region.
These states can be composed of sterile and active neutrino flavors
as described in the present section.

\section{Constraints from \Lv-processes. General pattern. }

Let us examine some generic features of those constraints on the Majorana
neutrino mass matrix which can be derived from \Lv-processes.

Majorana neutrino mass terms in Eq. \rf{mass-term} violate lepton number
conservation by two units $\Delta L = 2$ and, thus, can induce \Lv-processes
with $\Delta L = 2 n$. W-boson loops can be used to convert
$\Delta L \neq 0$ from the neutrino to the charged lepton sector.
Therefore \Lv-processes offer one of the most straightforward ways
to test the Majorana nature of neutrinos and extract information on
the neutrino mass matrix ${\cal M}^{(\nu)}$.

Note that at energies below the new physics thresholds only
$\Delta L = 2 n$ can be realized provided that baryon number is
conserved $\Delta B = 0$. This is a simple consequence of the
Lorentz invariance and the fact that the spinor SM fields are
represented only by leptons and quarks. To prevent $\Delta B \neq
0$ one has to contract the Lorentz indices of the external lepton
fields only with each other without involving quark fields. Thus
only \Lv-processes with even number of external leptons, i.e.
$\Delta L = 2 n$ processes, can proceed at these energies. This
means that any \Lv-process in this energy domain is related to the
Majorana neutrino mass receiving contributions from virtual
Majorana neutrino exchange. Certainly only $\Delta L =2$ processes
are of practical interest. The Majorana neutrino exchange
contribution to the rate of a $\Delta L =2$ process with two
external leptons $l_i l_j$ can be written schematically as
\ba{rate-Lv1}
\Gamma_{ij}= c \int\limits_{s_1^-}^{s_1^+} d s
\sum_{k} \left | \frac{U_{ik}U_{jk} m_{\nu k}}{s\pm m_{\nu k}^2}\right |^2
G(s/m_0^2) \ + \ ......
\ea
The function in the absolute value brackets originates from the \Lv
Majorana neutrino propagator $\lg 0|T(\nu(x) \nu^T(y))|0 \rg$.
Since the only source of \Lv in the neutrino sector is given by
the Majorana neutrino masses $m_{\nu k}$ the decay rate
\rf{rate-Lv1} vanishes when $m_{\nu k} = 0$. The ellipsis in this
equation denote terms whose explicit form is irrelevant for the
present general discussion. In Eq. \rf{rate-Lv1}  $G(z)$ is a
smooth positively definite smearing function which depends on the
particular process, $s_{1}^{\pm}$ are the limits of integration
determined by the masses of the external particles involved in the
process. The sign $+(-)$ in the denominator corresponds to the
t-(s-)channel neutrino exchange. For the s-channel contribution
the total neutrino width $\Gamma_{\nu}$ has to be taken into
account if masses in the resonant region of s-channel exchange
$s_1^-\leq m_{\nu}\leq s_1^+$ are considered. The non-zero
neutrino decay widths $\Gamma_{\nu k}$ can be introduced via the
substitution $m_{\nu k}\rightarrow m_{\nu k} - (i/2)\Gamma_{\nu k}$.

From the form of $\Gamma_{ij}$ one can infer that as a function on
neutrino masses $m_{\nu k}$ it has a maximal value
$\Gamma^{max}_{ij}$  for certain configuration of neutrino masses.
This observation leads to the conclusion that the sensitivity
$\Gamma^{Exp}$ of a concrete experiment searching for \Lv must
satisfy the condition $\Gamma^{Exp} \leq \Gamma^{max}_{ij}$
otherwise no information on neutrino contribution is derivable.
The experiment having passed this condition provides certain
constraints. Assume that neutrino fields can be divided into light
$\nu_i$ and heavy $N_i$ states with masses $m_{\nu i} <<
\sqrt{s_1^-}$  and $\sqrt{s_1^+} << M_N$ respectively. Then in
this ``light-heavy" neutrino scenario the Eq. \rf{rate-Lv1} can be
approximately rewritten as
\ba{approx-1} \Gamma_{ij} = \left|\lg m_{\nu}\rg_{ij}\right|^2
m_0^{-1} {\cal A}_{\nu}+ \left|\lg\frac{1}{M_N}\rg_{ij}\right|^2
m_0^3 {\cal A}_N \pm \mbox{Re}\left[\lg
m_{\nu}\rg_{ij}\lg\frac{1}{M_N}\rg_{ij}\right] m_0 {\cal A}_{\nu
N} \ea where the dimensionless coefficients ${\cal A}_{i}$ can be
obtained for a concrete process from the formula such as Eq.
\rf{rate-Lv1}. In the above equations $m_0\sim \sqrt{s_{1}^{\pm}}$
is a typical scale of the \Lv process under consideration.

The average masses in Eq. \rf{approx-1} are determined in the standard way
\ba{average}
\lg m_{\nu}\rg_{ij} = \sum_{k=light} U_{ik}U_{jk} m_{\nu k},\ \ \
\lg\frac{1}{M_N}\rg_{ij}  =  \sum_{k=heavy} \frac{U_{ik}U_{jk}}{M_{Nk}}.
\ea
Summation over light and heavy neutrinos implies masses
$m_{\nu k}<<\sqrt{s_1^-}$ and $M_{N k}>>\sqrt{s_1^+}$ respectively.

An experimental constraint on the rate of certain
\Lv process derived from its non-observation at
the experimental sensitivity $\Gamma^{Exp}$ can
be translated with the aid of Eq. \rf{approx-1} into the bounds:
\ba{constraints}
\Gamma_{ij}\leq\Gamma^{Exp}\longrightarrow
\left\{\begin{array}{c}
|\lg m_{\nu}\rg_{ij}| \leq Exp(\nu) \equiv
\sqrt{m_0 \Gamma^{Exp}/{\cal A}_{\nu}},\\
|\lg\frac{1}{M_N}\rg_{ij}| \leq Exp(N) \equiv
\sqrt{\Gamma^{Exp}/(m_0^3 {\cal A}_{\nu})}.
\end{array}\right.
\ea
However this is only possible if
the experimental sensitivity satisfies
the consistency conditions
\ba{consist}
Exp({\nu})<< \sqrt{s_1^{-}}\sim m_{0}, \
Exp({N})^{-1} >>  \sqrt{s_1^{+}}\sim m_{0}.
\ea 
Otherwise experimental data can not be translated 
into the constraints  \rf{constraints} as it was done, for instance, in Refs.
\cite{Kdec4,NMF,HERA}. If the consistency conditions \rf{consist}
are not satisfied one has to use the initial formula
\rf{rate-Lv1}.

The following remark is in order. If all the  neutrino states are light
satisfying $m_{\nu k}<<\sqrt{s_1}$ then the following relation takes place
\ba{entry-1}
\lg m_{\nu}\rg_{ij} = {\cal M}^{(\nu)}_{ij}
\ea
This relation is not true if there are heavy neutrino states
with masses not satisfying the condition $m_{\nu}<<\sqrt{s_1^-}$.
According to Eq. \rf{average} they do not contribute
to  $\lg m_{\nu}\rg_{ij}$ measured in some \Lv process.
Therefore a concrete \Lv process can give direct information on the entry
$ {\cal M}^{(\nu)}_{ij}$ of the Majorana neutrino mass matrix only under
the assumption that all the neutrino masses are small compared to a typical
scale of this process $m_{\nu k}<<m_0\sim \sqrt{s_{1}^{\pm}}$, or assuming that
the heavy states are sterile. From this point of view \Lv processes with
larger typical scales $m_0$ are preferable.

In the subsequent sections we will study concrete \Lv processes
from the viewpoint of their ability to probe neutrino properties.
We will consider the conventional neutrino spectrum with three
light neutrinos plus a number of kinematically unattainable heavy
states as well as a model with  additional intermediate mass
neutrinos in the hundred MeV domain.

\section{K-meson neutrinoless double muon decay }

Here we shortly outline the theoretical framework
for the $\kd$ decay. Recently this \Lv process attracted
attention \cite{LS:2000,GDKS:07.2000} as a possible probe
of the neutrino sector complimentary to other known processes.

In the SM extension with Majorana
neutrinos there are two lowest order diagrams, shown in Fig.1,
which contribute to the $\kd$ decay.
We concentrate on the s-channel neutrino exchange diagram in Fig.~1(a)
which plays a central role in our analysis.
The t-channel diagram in Fig.~1(b) requires
in general a detailed hadronic structure calculation. In
Ref. \cite{Kdec2} this diagram was evaluated in the Bethe-Salpeter
approach and shown to be an order of magnitude smaller than the
diagram in Fig.~1(a), for light and intermediate mass neutrinos.

The contribution from the factorizable s-channel diagram in
Fig.~1(a) can be calculated in a straightforward way, without
referring to any hadronic structure model. A final result for the
$\kd$ decay rate is given by \cite{GDKS:07.2000}
\ba{rate-Lv}
&&\Gamma(\kd)= c \int\limits_{s_1^-}^{s_1^+} d s_1
\left |\sum_{k} \frac{U_{\mu k}^2 m_{\nu k}}{s_1 - m_{\nu k}^2}\right |^2
G(\frac{s_1}{m_{_K}^2}) + \\ \nn
&&2 \frac{c}{m_{_K}^{2}}{\rm Re}\sum_{k,n}[\int\limits_{s_1^-}^{s_1^+} d s_1
\frac{U_{\mu k}^2 m_{\nu k}}{s_1 - m_{\nu k}^2}
\int\limits_{s_2^-}^{s_2^+} d s_2
\left(\frac{U_{\mu n}^2 m_{\nu n}}{s_2 - m_{\nu n}^2}\right)^*
H(\frac{s_1}{m_{_K}^2},\frac{s_2}{m_{_K}^2})].
\ea
The unitary mixing matrix $U_{ij}$ relates $\nu'_i = U_{ij}\nu_j$
weak $\nu'$ and mass $\nu$ neutrino eigenstates.
The numerical constant in Eq. \rf{rate-Lv} is
\ba{const1}
c = (G_F^4/32) (\pi)^{-3} f^2_{\pi}  f^2_{_K} m^5_{_K}
|V_{ud}|^2|V_{us}|^2,
\ea
where $f_{_K} = 1.28 ~f_{_\pi}$, $f_{_\pi} = 0.668~ m_{\pi}$
and $m_K = 494$ MeV is the K-meson mass.
The functions $G(z)$ and $H(z_1,z_2)$ in Eq. \rf{rate-Lv}
after the phase space integration can be written in an explicit
algebraic form
\ba{G-funct}
G(z) &=& \frac{\phi(z)}{z^{2}}
\left[h_{+-}(z)h_{--}(z)-
x_{\pi}^2h_{-+}(z)\right]
\left[x_{\mu}^2 +z -(x_{\mu}^2-z)^2\right]\\ \nn
H(z_1,z_2)&=& h_{--}(z_1)h_{--}(z_2) +
x_{\pi}^2[r_{+}(z_1z_2) - x_{\mu}^2t(z_1,z_2,1)]-
r_{-}(z_1z_2)t(z_1,z_2,x_{\mu}).
\ea
Here we defined $x_{i} = m_i/m_{_K}$ and introduced the functions
\ba{def10}
&&h_{\pm\pm}(z)=z\pm x^2_{\pi}\pm x^2_{\mu}, \ \ \
r_{\pm}(z_1z_2)= z_1z_2 -x_{\pi}^2 \pm x_{\mu}^4,\\ \nn
&&t(z_1,z_2,z_3)= z_1+z_2-2 z_3^2, \ \ \
\phi(z)=
\lambda^{1/2}(1,x_{\mu}^2,z)\lambda^{1/2}(z,x_{\mu}^2,x_{\pi}^2)
\ea
with $\lambda(x,y,z) = x^2+y^2+z^2-2xy -2yz- 2xz$.

The integration limits in Eq. \rf{rate-Lv} are
\ba{lim-int}
&&s_1^- = m_{_K}^2(x_{\pi} + x_{\mu})^2, \ \ \ \
s_1^+ = m_{_K}^2(1 - x_{\mu})^2,\\ \nn
&&s_2^{\pm} = \frac{m_{_K}^2}{2y} \left[2y(1+x_{\mu}^2)-
(1+y-x_{\mu}^2)h_{-+}(y)\pm \phi(y) \right]
\ea
with $y=s_1/m_{_K}^2$.

Assuming that neutrinos can be separated into
light $\nu_k$ and heavy $N_k$ states, with masses
$m_{\nu i} << \sqrt{s_1^-}$  and $\sqrt{s_1^+} << M_{N k}$
we can rewrite Eq. \rf{rate-Lv} in the approximate form
\rf{approx-1} with the  dimensionless coefficients
\ba{A-k}
{\cal A}_{\nu} = 4.0\times 10^{-31}, \
{\cal A}_{N} = 7.0\times 10^{-32},  \
{\cal A}_{\nu N} = 1.7\times 10^{-31}.
\ea
With these numbers we can estimate the current upper bound
on the $\kd$ decay rate from the experimental data on other processes.

\section{``Light-Heavy" neutrino scenario}

Here we assume that all the neutrino mass eigenstates can
be divided into very light  $\nu_i$ and very heavy $N_i$ states with
masses, respectively, much smaller and much larger than the characteristic
energy scale $m_0$ of the studied  \Lv process. Let us consider in this ``Light-Heavy"
neutrino scenario several typical examples of \Lv processes and estimate
their ability to constraint the average masses $\lg m_{\nu}\rg_{ij},\  \lg 1/M_N\rg_{ij}$
as well as their possible rates.

At present the highest experimental sensitivity to the Majorana neutrino
contribution has been achieved in neutrinoless double beta decay ($\znbb$)
$(A,Z)\rightarrow (A,Z+2) + 2 e^{-}$. A typical scale of this process is set by
the nucleon Fermi momentum $m_0\sim p_{_F}\approx 100$MeV.
The current constraints
from $\znbb$ decay are \cite{znbb-exp,znbb-rev2}
\ba{znbb-constr}
|\lg m_{\nu}\rg_{ee}| \leq 0.2-0.6\ \mbox{eV},\ \ \ \
|\lg\frac{1}{M_N}\rg_{ee}|\  \lsim \ \left(9.0 \cdot 10^{7}\mbox{GeV}\right)^{-1}.
\ea
The uncertainty relates to the uncertainties in the nuclear matrix elements 
and treatment of the background conditions. The first constraint in Eq. \rf{znbb-constr}, 
assuming that all the neutrinos are much lighter than 100 MeV, provides a direct constraint on 
the $ {\cal M}^{(\nu)}_{ee}$
entry of the Majorana neutrino mass matrix. Evidently these constraints satisfy 
the consistency conditions \rf{consist}.

Experiments searching for the other  \Lv processes
have not yet reached enough sensitivity to establish meaningful
constraints directly on the neutrino mass matrix elements.

For instance, experiments on the muon to positron
nuclear conversion
$\mu^- + (A,Z)\rightarrow e^+ + (A, Z-2)$ in $^{48}$Ti
give at current sensitivity the following upper bound on
the branching ratio
\cite{mu-e-exp}
\ba{mu-e}
{\rm R}(\mu^-\rightarrow e^+) =
\frac{\Gamma({\rm Ti} +
\mu^-\rightarrow {\rm Ca} + e^+)}{\Gamma({\rm Ti} + \mu^-
\rightarrow {\rm Sc} + \nu_{\mu})} \leq 1.7 \cdot 10^{-12} \ \ \ (\mbox{90\%CL})
\ \ \
\ea
Assuming that all the neutrinos are much lighter that the typical energy
scale $m_0\sim m_{\mu}=105$MeV of this reaction one finds the bound
\ba{mu-e-const}
|\lg m_{\nu}\rg_{\mu e}| \leq 17(80)\mbox{MeV}
\ea
for the proton pairs of the final nucleus
in the singlet(triplet) state\cite{DKT}. This constraints
are marginal from the viewpoint of the consistency condition \rf{consist}.

Direct searches for $\kd$ decay by E865 experiment
at BNL \cite{E865} give
\ba{kdec-lim}
{\cal R}_{\mu\mu} = \frac{\Gamma(\kd)}{\Gamma(K^+\rightarrow all)} \leq 3.0\times 10^{-9}
\ \ \ \left(\mbox{ 90\%CL}\right).
\ea
Applying to this case the approximate formula \rf{approx-1} one gets the limit
\ba{erron}
|\lg m_{\nu}\rg_{\mu\mu}| \leq 500\mbox{GeV}
\ea
which makes no sense because it does not satisfy the condition
\rf{consist} with the typical energy scale of this process $m_0\sim m_{_K}=494$MeV.
Thus the approximate formula \rf{approx-1} is not applicable to the present
experimental situation for the $\kd$ searches. A similar picture holds for all known
\Lv processes searched for in various experiments (for other examples see  Refs. \cite{lnv}).

Viewing these processes from the stand point of neutrino observations
one finds that except for the $\znbb$-decay they have very small rates in
the ``Light-Heavy" neutrino scenario.

Atmospheric and solar neutrino oscillation data
demonstrate $\delta m^2<<(1 eV)^2$ suggesting that
all the neutrino mass eigenstates are approximately degenerate
at the 1 eV scale \cite{barg1}. This observation in combination
with the tritium beta decay endpoint allows one to set upper bounds
on masses of all the three neutrinos \cite{barg1}
$m_{e,\mu,\tau}\leq 3$eV. Thus in the three neutrino scenario
one derives
\ba{3-neutrino}
|\lg m_{\nu}\rg_{ij}|  \leq 9\mbox{eV}\ \ \ \ \mbox{for} \ \ \ \ i,j = e, \mu, \tau.
\ea
This is much lower than the existing constraints on this quantity
from \Lv processes, except for $\znbb$-decay which gives a significantly more
stringent upper bound \rf{znbb-constr}. With the constrain \rf{3-neutrino}
we can predict the rate of various \Lv processes.

Let us substitute the upper bound \rf{3-neutrino} into the formula
\rf{approx-1} written for the $\kd$ decay with coefficients given in Eq. \rf{A-k}.
This gives rise to the following extremely small branching ratio
\ba{limit11}
{\cal R}_{\mu\mu} =
\frac{\Gamma(\kd)}{\Gamma({\rm K}^+\rightarrow all)}
\leq 3.0 \times 10^{-30}  \qquad
{\rm (3\ light\ neutrino\ scenario)}.
\ea
Assume there exist in addition heavy neutrinos N with
the masses in the GeV region. Using the current
LEP limit on heavy unstable neutral leptons $M_N\geq 54.4$ GeV \cite{LEP}, we get
\ba{lim-N}
|\lg M_N^{-1}\rg_{\mu\mu}| \leq n \left(54.4 \mbox{ GeV}\right)^{-1},
\ea
where $n$ is the number of heavy neutrinos.

This limit being substituted in Eq.~\rf{approx-1} together with
the limit \rf{3-neutrino} results in the upper bound
\ba{limit12}
{\cal R}_{\mu\mu} \leq 1.0\times 10^{-19} \qquad {\rm
(3\ light +  1\ heavy\ neutrino\ scenario)}.
\ea

Comparison of the theoretical predictions in Eqs.\ \rf{limit11}, \rf{limit12}
with  the experimental bound in Eq. \rf{kdec-lim} clearly shows that both cases
are far from being ever detected. A similar conclusion is true for the other
\Lv processes except $\znbb$ decay.

On the other hand experimental observation of these processes
at larger rates would indicate some new physics beyond the SM, or,
as we will see for the case of the $\kd$ decay, the presence of an extra
neutrino state $\nu_j$ with mass in the hundred MeV domain.
As we discussed in section 2, the extra massive neutrino states
$\nu_j$ can appear as a result of mixing of the three active
neutrinos with certain number of sterile neutrinos. These massive
neutrinos are at present searched for in many experiments \cite{PDG}. The
$\nu_j$ states would manifest themselves as peaks in differential
rates of various processes, and can give rise to significant
enhancement of the total rate if their masses lie in an
appropriate region.

\section{Hundred MeV neutrinos in $\kd$-decay}

Assume there exists a massive Majorana neutrino $\nu_j$
with the mass $m_j$
\ba{domain}
\sqrt{s_1^-} \approx 245 \mbox{ MeV} \leq m_{j} \leq \sqrt{s_1^+} \approx 389
\mbox{ MeV}.
\ea
In this mass range the s-channel neutrino exchange diagram in Fig.~1(a) absolutely
dominates over the t-channel diagram in Fig.~1(b), independently of hadronic
structure. Here the diagram in Fig.~1(a) blows up because the integrand of the first
term in Eq.~\rf{rate-Lv} has a non-integrable singularity at $s =m_j^2$. Therefore, in this
resonant domain the total $\nu_j$-neutrino decay width
$\Gamma_{\nu j}$ has to be taken into account. This can be done by
the substitution $m_{j}\rightarrow m_{j} - (i/2)\Gamma_{\nu j}$.

The total decay width $\Gamma_{\nu j}$ of the Majorana neutrino
$\nu_j$ with mass in the resonant domain \rf{domain} receives contributions
from the following decay modes:
\ba{channels}
\nu_j\longrightarrow
\left\{\begin{array}{l}
e^+\pi^-,\ e^-\pi^+,\ \mu^+\pi^-,\ \mu^-\pi^+,\\
e^+e^-\nu_e^c,\ e^+\mu^-\nu_{\mu}^c,\ \mu^+e^-\nu_e^c,\ \mu^+\mu^-\nu_{\mu}^c\\
e^-e^+\nu_e,\ e^-\mu^+\nu_{\mu},\ \mu^-e^+\nu_e,\ \mu^-\mu^+\nu_{\mu}.
\end{array}\right. ,
\ea
Since $\nu_j\equiv \nu_j^c$ it can decay in both
$\nu_j\rightarrow l^{-} X(\Delta L=0)$ and
$\nu_j\rightarrow l^{+} X^c(\Delta L=2)$ channels.
Calculating partial decay rates we obtain \cite{GDKS:07.2000}
\ba{dec-width-4}
\Gamma(\nu_j\rightarrow l \pi)= |U_{lj}|^2 \frac{G_F^2}{4\pi}f_{\pi}^2
m_j^3 F(y_l,y_{\pi})\equiv |U_{lj}|^2 \Gamma_2^{(l)},\\
\Gamma(\nu_j\rightarrow l_1l_2\nu )=
|U_{l_1 j}|^2 \frac{G_F^2}{192\pi^3} m_j^5 H(y_{l1},y_{l2})
\equiv |U_{l_1 j}|^2 \Gamma_3^{l_1l_2},
\ea
where $y_i= m_i/m_j$ and
\ba{kin-fun}
&&F(x,y)= \lambda^{1/2}(1,x^2,y^2)
[(1+x^2)(1+x^2-y^2) - 4 x^2],\\
&&H(x,y)= 12 \int\limits_{z_1}^{z_2} \frac{dz}{z}
(z-y^2)(1+x^2-z)
\lambda^{1/2}(1,z,x^2)\lambda^{1/2}(0,y^2,z).
\ea
The integration limits are $z_1 = y_{l_2}^2,\ z_2=(1-y_{l_1})^2$ and
$F(0,0)=H(0,0)=1$.
Summing up all the decay modes in \rf{channels} one gets for the total
$\nu_j$ width
\ba{total-4}
\nn
&&\Gamma_{\nu j} =
2 |U_{\mu j}|^2(\Gamma_2^{(\mu)} + \Gamma_3^{(\mu e)}+
\Gamma_3^{(\mu\mu)})+
2 |U_{ej}|^2(\Gamma_2^{(e)} + \Gamma_3^{(ee)}+ \Gamma_3^{(e\mu)})\equiv\\
&&\equiv |U_{\mu j}|^2\Gamma_{\nu}^{(\mu)} + |U_{ej}|^2\Gamma_{\nu}^{(e)}.
\ea
In the resonant domain \rf{domain}  $\Gamma_{\nu j}$ reaches
its maximum value at $m_j = \sqrt{s_1^+}$. Assuming for the moment
$|U_{\mu j}|=|U_{ej}|=1$, we estimate this maximum value to be
$\Gamma_{\nu j}\approx 4.7\times 10^{-10}$ MeV.
Since $\Gamma_{\nu j}$ is so small in the resonant domain
\rf{domain} the neutrino propagator in the first term of Eq.~\rf{rate-Lv}
has a very sharp maximum at $s=m_{j}^2$. The second term, being finite
in the limit $\Gamma_{\nu j} =0$, can be neglected
in the considered case. Thus, with a good
precision we obtain from Eq.~\rf{rate-Lv} \cite{GDKS:07.2000}
\ba{estim3}
\Gamma^{res}(\kd) \approx
c \pi G(z_0)\frac{m_j |U_{\mu j}|^4}
{|U_{\mu j}|^2\Gamma_{\nu}^{(\mu)} + |U_{ej}|^2\Gamma_{\nu}^{(e)}}
\ea
with $z_0 = (m_j/m_K)^2$.
This equation allows one to derive, from the experimental
bound of Eq.~\rf{kdec-lim}, constraints on the $\nu_j$ neutrino mass $ m_{j}$
and the mixing matrix elements
$U_{\mu j}, U_{e j}$ in a form of a 3-dimensional exclusion plot.
However one may reasonably assume that $|U_{\mu j}|\sim |U_{ej}|$.
Then from the experimental bound \rf{kdec-lim} we derive a
2-dimensional $m_j-|U_{\mu j}|^2$ exclusion plot given in Fig.~2.
For comparison we also present in Fig.~2 the existing bounds taken
from \cite{LEP1}.
As shown in the figure, the experimental data on
the $\kd$ decay exclude a region unrestricted by the other
experiments. The constraints can be summarized as
\ba{most}
|U_{\mu j}|^2 \leq (5.6\pm 1)\times 10^{-9}\ \ \ \ \
\mbox{for}\ \ \ 245\mbox{ Mev}\leq m_j \leq 385\mbox{ MeV},
\ea
The best limit $|U_{\mu j}|^2 \leq 4.6\times 10^{-9}$ is achieved
at $m_j \approx 300$ MeV. Note that these limits are compatible with our
assumption that $|U_{\mu j}|\sim|U_{ej}|$ since in this mass domain,
typically $|U_{ej}|^2\leq 10^{-9}$ \cite{PDG}.

The constraints from $\kd$ in Fig. 2 and Eq. \rf{most} can be
significantly improved in the near future by the experiments E949 at
BNL and E950 at FNAL \cite{FNAL}.
It is important to notice that in the resonant
domain we have $\Gamma^{res}(\kd)\sim |U|^2$,
while outside $\Gamma(\kd)\sim |U|^4$. Thus in the resonant mass
domain the $\kd$ decay has a significantly better sensitivity to
the neutrino mixing matrix element. In forthcoming experiments
the upper bound on the ratio in Eq.~\rf{kdec-lim} can be improved by
two orders of magnitude or even more.
Then this experimental bound could be translated to
the limit $|U_{\mu j}|^2 < 10^{-11}$ and stronger.

\section{Hundred MeV neutrinos in astrophysics and cosmology}

It is well known that massive neutrinos may have
important cosmological and astrophysical implications.
They are expected to contribute to the mass density of
the universe, participate in cosmic structure formation,
big-bang nucleosynthesis, supernova explosions, imprint themselves
in the cosmic microwave background etc.\ (for a review see, for instance,
Ref.  \cite{Raffelt}). This implies certain constrains on the neutrino masses
and mixings.
Currently, for massive neutrinos in the mass region
~\rf{domain}, the only available cosmological
constraints arise from the mass density of the
universe and cosmic structure formation.

The contribution of stable massive neutrinos to the mass density
of the universe is described by the ``Lee-Weinberg" $\Omega_{\nu}
h^2-m_{\nu}$ curve. From the requirement that the universe is not
``overclosed" this leads to the two well know solutions $m_{\nu}\
\leq 40$eV and $m_{\nu} \ \gsim 10$GeV which seem to exclude the
domain Eq. \rf{domain}. However for the unstable neutrinos the
situation is different.  They may decay early to light particles
and, therefore, their total energy can be significantly
``redshifted" down to the ``overclosing" limit. Constraints on the
neutrino life times $\tau_{\nu_j}$ and masses $m_j$ in this
scenario were found in Ref. \cite{DM}. In the mass region
\rf{domain} we have an order of magnitude estimate
\ba{DM}
\tau_{\nu_j}< (\sim 10^{14}) \mbox{sec} \ \ \ \ \ \mbox{Mass Density limit}
\ea

Decaying massive neutrinos may also have specific impact on
the cosmic structure formation introducing new stages
in the evolution of the universe. After they decay into light relativistic
particles the universe returns for a while from the matter to the radiation
domination phase. This may change the resulting density fluctuation spectrum since
the primordial fluctuations grow due to gravitation instability during the matter
dominated stages.
Comparison with observations leads to an upper bound on
the neutrino life time \cite{StructForm}. In the mass region \rf{domain}
one finds roughly
\ba{SF}
 \tau_{\nu_j}< (\sim 10^{7})\mbox{sec} \ \ \  \mbox{Structure Formation limit}
\ea

On the other hand, on the basis
of formula \rf{total-4}, assuming $|U_{\mu j}|^2 \sim
|U_{ej}|^2 \leq 4.6\times 10^{-9}$ as in Eq.~\rf{most}, we find
conservatively
\ba{theor}
10^{-2} \mbox{sec} < \tau_{\nu_j}, \ \ \ \ \mbox{Theoretical limit.}
\ea
Thus massive neutrinos with masses in the interval \rf{domain} are not yet
excluded by the known cosmological constraints \rf{DM}, \rf{SF}
and there remains a wide open interval of allowed mixing matrix elements:
\ba{open}
 (\sim 10^{-18}) < |U_{\mu j}|^2,\ |U_{ej}|^2 < (\sim 10^{-9}).
\ea

Big-bang nucleosynthesis and the SN 1987A neutrino signal may presumably
lead to much more restrictive constraints \cite{Dolgov}. Unfortunately, as yet
the analysis \cite{Dolgov} of these constraints does not involve
the mass region \rf{domain}. It may happen that these constraints,
in combination with our constrains in Eq.~\rf{most}, close the
window for neutrinos with masses in the interval \rf{domain}. Then
the only physics left to be studied using the $\kd$ searches would
be physics beyond the SM other than neutrino issues. Nevertheless,
significant model dependence of all the cosmological constraints
should be carefully considered before such a determining
conclusion is finally drawn.

\section{Conclusion}

We analyzed some generic properties of $\Delta L=2$
lepton-number violating processes and the constraints
derivable from them on neutrino masses and mixing matrix elements.
We discussed consistency conditions for experimental bounds
when these bounds can be translated into the upper limits
on the average neutrino mass $\lg m_{\nu}\rg$ or the inverse
average mass $\lg 1/M_N\rg$. We found that excepting the neutrinoless
double decay other $\Delta L =2$ processes are unable to provide us with
sensible constraints on these quantities.
Using the neutrino oscillation data, the tritium beta decay and the LEP searches
for the heavy neutral lepton we estimated constraints on
their rates in the scenario with three light and several heavy neutrinos.
Typical values of these rates are far from being reached experimentally
in the near future.

We studied the potential of the K-meson neutrinoless double muon
decay as a probe of Majorana neutrino masses and mixings. We found that
this process is very sensitive to the hundred MeV neutrinos $\nu_j$ in the resonant
mass  range \rf{domain}. We analyzed the contribution of these neutrinos
to the $\kd$ decay rate and derived stringent upper limits on Majorana
neutrino mixing matrix element $|U_{\mu j}|^2$ from current experimental
data.
In Fig.~2 we presented these limits in the form of
a 2-dimensional exclusion plot, and compared them with existing limits.
The $\kd$ decay excludes a domain previously unrestricted experimentally.
We stressed that the known astrophysical and cosmological constraints
do not yet exclude hundred MeV neutrinos satisfying these $\kd$ constraints.

Finally, we notice that the decay $\kd$ can in principle probe
lepton number violating interactions beyond the standard model.
However, according to recent studies \cite{LS:2000,Kdec-SUSY},
supersymmetric interactions both with and without R-parity
conservation seem to be beyond the reach of the K-decay experiments.

\vskip10mm
\centerline{\bf Acknowledgments}
We are grateful to F. Vissani for useful comments and remarks. 
This work was supported in part by Fondecyt (Chile) under grants
1990806, 1000717, 1980150 and 8000017, and by a C\'atedra
Presidencial (Chile). \bigskip
%

\newpage
\begin{figure}[h!]
\vspace{-2 cm}
\hspace{-0.5 cm}

\mbox{ \epsfxsize=16 cm\epsffile{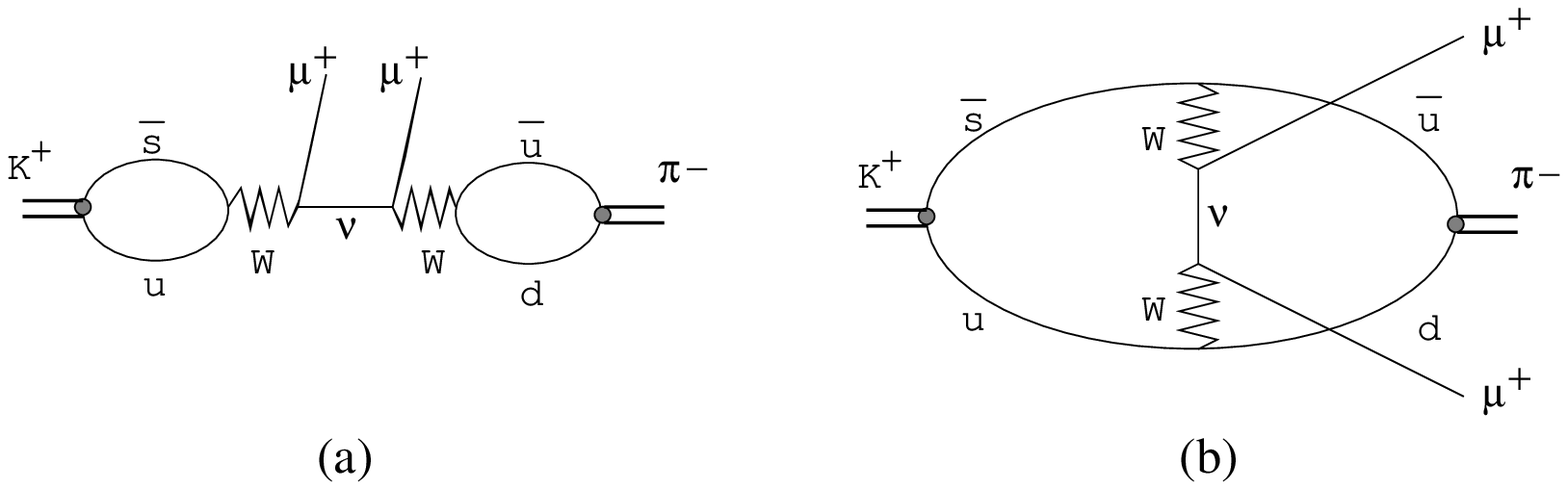}}
\caption{The lowest order diagrams contributing to $\kd$ decay.}
\end{figure}
\begin{figure}[h!]
\hspace{-0.5 cm}
\mbox{\epsfxsize=14 cm\epsffile{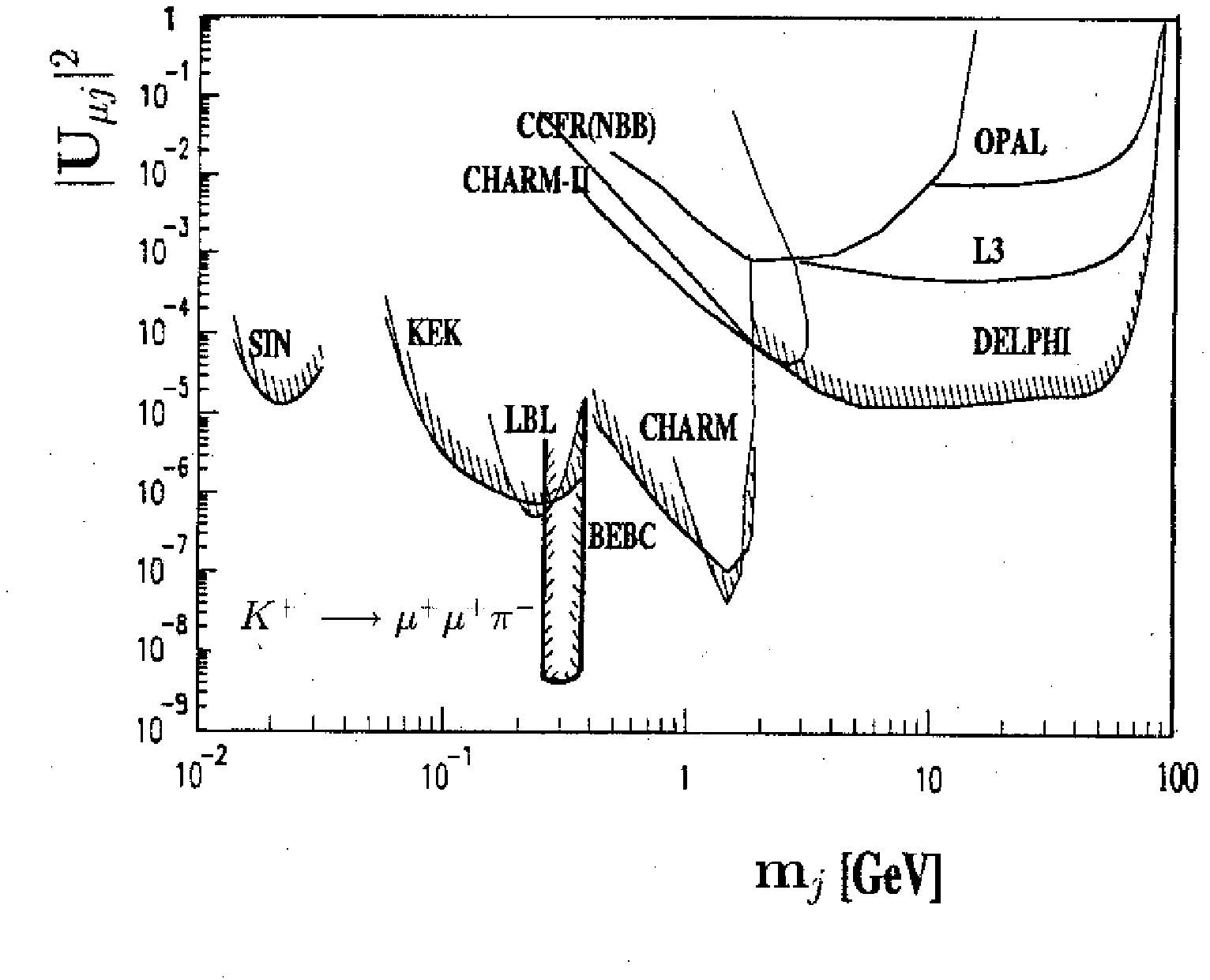}}
\caption{Exclusion plots in the plane $|U_{\mu j}|^2-m_j$.
Here $U_{\mu j}$ and $m_j$ are the heavy neutrino $\nu_j$
mixing matrix element to $\nu_{\mu}$ and its mass respectively.
Domains above the curves are excluded by various experiments
according to the recent update in Ref. \protect\cite{LEP}.
Region excluded by $\kd$ decay  \protect\cite{GDKS:07.2000}
covers the interval $249$MeV$\leq m_j\leq 385$MeV and extends down to
$|U_{\mu j}|^2\leq 4.6\times 10^{-9}$.}
\end{figure}
\end{document}